\documentclass{aa}
\input psfig.tex

\newcommand{\ms}{M$_{\odot}$}
\newcommand{\zs}{Z$_{\odot}$}
\newcommand{\dkp}{dex kpc$^{-1}$}

\begin{document}

\thesaurus{ 04(08.01.1; 02.14.1; 10.01.1; 10.05.1) }
\title{Abundance gradients and their evolution in the Milky Way disk}

\author{J.L. Hou \inst{1,2,3,4},
     N. Prantzos \inst{2} and
     S. Boissier \inst{2}       } 

\institute{Shanghai Astronomical Observatory, CAS, Shanghai, 200030, P.R. China
(hjlyx@center.shao.ac.cn) 
\and Institut d'Astrophysique de Paris, 98bis, Bd. Arago, 75014, Paris, France
(prantzos@iap.fr)
\and National Astronomical Observatories, CAS, P.R. China 
\and Joint Lab of Optical Astronomy, CAS, P.R. China 
}
\date {}
\maketitle 


\begin{abstract}

Based on a simple, but fairly successful, model of the chemical evolution
of the Milky Way disk, we study the evolution of the abundances of
the elements He, C, N, O, Ne, Mg, Al, Si, S, Ar and Fe. We use
metallicity dependent yields for massive stars with and without mass
loss. We find that most observed abundance profiles are correctly 
reproduced  by massive star yields, but C and N require supplementary 
sources. We argue that massive, mass losing stars can totally account 
for the abundance profile of C, while intermediate mass stars are the 
main source of N; in both cases, some conflict with corresponding data
on extragalactic HII regions arises, at least if current observations
in the Galaxy are taken at face value. The observed behaviour of Al
is marginally compatible with current massive star yields, which probably
overestimate the ``odd-even'' effect. We also find that the adopted
``inside-out'' formation scheme for the Milky Way disk produce abundance
profiles steeper in the past. The corresponding abundance scatter
is smaller in the inner disk than in the outer regions for a given
interval of Galactic age.

\keywords{Stars:abundances - Stars:yields - Galaxy:gradient - Galaxy:evolution } 

\end{abstract}

\section{Introduction}

Abundance gradients constitute one of the most important 
observational constraints for  models of the 
evolution of the Milky Way disk.
The existence of such gradients is now well established, through
radio and optical observations of HII regions, stars 
and planetary nebulae (see Henry and Worthey 
1999 for a review of the abundance profiles in the Milky Way 
as well as in external galaxies). 
An average gradient of dlog(X/H)/dR$\sim-$0.06 \dkp \ is 
observed in the Milky Way for O, S, Ne and Ar, while  the
gradients for C and N are, perhaps, slightly larger (Smartt 2000).
Observations of open clusters 
support the above picture (Friel 1995, 1999; Carraro et al. 1998), 
albeit with large uncertainties.

The existence of those gradients offers the opportunity to test theories
of disk evolution and stellar nucleosynthesis. Indeed, the
magnitude of the observed gradients in the Milky Way disk is rather
large: metal abundances in the inner disk are larger than those
in the outer disk by a factor of $\sim$10. This suggests that the 
role of the Galactic bar in inducing large
scale radial mixing and therefore flatenning the abundance profile (e.g.
Friedli and Benz 1995) has been rather limited; alternatively, the
Galactic bar is too young ($<$1 Gyr) 
to have brought any important modifications
to the gaseous and abundance profiles of the disk.
In any case, the magnitude of the observed gradients puts important 
constraints on disk models invoking radial inflows (see e.g.
Portinari and Chiosi 2000 and references therein). Even in the
case of simpler models (with no radial inflows), the abundance gradients
constrain the various parameters, like the local timescales of
star formation and infall (e.g. Matteucci and Francois 1989,
Prantzos and Aubert 1995, Moll\`a et al. 1997) or any
variation of the stellar Initial Mass Function properties with
metallicity (e.g. Chiappini et al. 2000). 
On the other hand, profiles of abundance ratios across the disk
constrain the nature of various elements and isotopes (i.e.
``secondaries'' vs. ``primaries'', ``odd-Z'' vs. ``even-Z'') and their
nucleosynthesis  sites (e.g. Prantzos et al. 1996).

A large number of chemical evolution models has been developped
in the 90ies, aiming to explore one or more of the above issues,
either with radial inflows (e.g. G\"otz and K\"oppen 1992,
Chamcham and Tayler 1994,
Tsujimoto et al. 1995, Firmani et al. 1996,
Thon and Meusinger 1998, Portinari and Chiosi 2000)
or without radial flows (Ferrini et al. 1994, Prantzos and Aubert 1995,
Giovagnoli and Tosi 1995, 
Carigi 1996, Chiapinni et al. 1997, Boissier and Prantzos 1999,
Chang et al. 1999). Some promising chemodynamical evolutionary models
for the Milky Way have also been developed by a few groups (Steinmetz and
M\"uller 1994, Samland et al. 1997, Berczik 1999).
For the simplest of those models (namely those with no radial flows), some
convergence has been recently reached between the various groups, at
least concerning the basic ingredients (e.g. Tosi 2000):
i) necessity of substantial infall and of radially varying timescales
for the infall and the star formation,  and  ii) no need for
varying IMF or strong galactic winds.

Despite this agreement, some important differences still exist between
the various models. The most important concerns the history of the
abundance profiles: were they steeper or flatter in the past? The
former is suggested by models of e.g. Prantzos and Aubert (1995), 
Moll\`a et al. (1997),
Allen et al. (1998), Boissier and Prantzos (1999), while the latter
is supported by models of Tosi (1988) and Chiappini et al. (1997).
The situation is not settled observationally either.
Estimated ages of open clusters and planetary nebulae of various types
span a large fraction of the age of the Galaxy. Observations of the
abundances of those objects across the Milky Way disk could,
in principle, provide some information on the past history of the
abundance gradients.
In practice, however, observational uncertainties are too large
at present to allow conclusions on that matter (e.g. Carraro et al. 1998, 
Maciel 1999).

Several observational studies on the abundance profiles of the Milky Way 
disk were performed in the late 90ies. They concerned a variety of elements 
(He, C, N, O, Ne, Mg, Al, Si, S, Ar and Fe), observed in HII regions 
(Simpson et al. 1995, Rudolph et al. 1997, Afflerbach et al. 1997, 
Deharveng et al. 2000), B-stars (Smartt and Rolleston 1997, Gummersbach 
et al. 1998) and planetary nebulae (Maciel and K\"oppen 1994, Maciel and 
Quireza 1999). Although they leave the question of the history of abundance
profiles unsettled, these studies allow for some important tests of stellar
nucleosynthesis theories, since they provide absolute abundances,
and abundance ratios along the Milky Way disk.

In a previous work (Boissier and Prantzos 1999, herefater BP99) we developed 
a simple model for the chemical and spectrophotometric evolution of the 
Milky Way disk. With few assumptions (basically concerning the
radial variation of the infall and star formation timescales),  
the model reproduces all the major obsevational constraints for the Galaxy:
total amounts of gas and stars, supernovae rates, radial profiles
of gas, stars, star formation rate and oxygen abundance, luminosities
and scalelengths in various wavelength bands. The success
of the model does not guarantee, of course, its correctness or its
uniqueness. However, it offers a sound basis
for a more detailed exploration of issues related to the chemical evolution
of the Milky Way. Two such issues are explored here:

i) how successful are currently available stellar yields in reproducing
the wealth of recently observed abundance profiles in the Galaxy? are
Wolf-Rayet  (massive, mass losing) stars or intermediate mass stars (IMS)
important in shaping the abundance profiles, and for which elements?

ii) is the evolution of abundance gradients predicted by the model
compatible with observations of available tracers? are there any
new, potentially testable, predictions of the model in that respect?

The plan of the paper is as follows: 
In Sect. 2, we present briefly the basic ingredients and the underlying 
assumptions of the BP99 model;  in particulrar, we present in some detail
the metallicity dependent yields of Woosley  and Weaver (1995) and 
Maeder (1992), on which much of our work is based. 
Model results for abundance gradients of  
He, C, N, O, Ne, Mg, Al, Si, S, Ar and Fe
and their evolution are given in Sect. 3. 
Sect. 4 constitutes the main body of the paper. 
In sect. 4.1 we present a compilation of observational 
data from HII regions,  B-stars and planetary nebulae in the Milky Way 
disk. Detailed comparison between model predictions and current observed disk 
profiles is made in Sect. 4.2. In Sect. 4.3, we further 
compare the behaviour of model profiles of abundance ratios to observations
and draw some conclusions on the adopted yields. Then, based on 
the abundance data for objects of different ages we discuss the 
evolution of abundance gradients for elements O, Ne, S and Ar (Sect. 4.4). 
The results  are summarized in Sect. 5.

\section{Model}

The adopted model for the chemical evolution of the Milky Way disk is described 
in detail in BP99. In Sect. 2.1 we briefly recall the main features of the model.
In Sect. 2.2 we present in some more detail the only novel ingredient with 
respect to BP99, namely the metallicity dependent yields of Woosley and Weaver 
(1995, hereafter WW95) for intermediate mass elements and of Maeder 
(1992, hereafter M92) for He, C, N, O (in BP99 only yields for stars of 
solar metallicity are used).

\subsection{Description of the model}

The galactic disk is considered as an ensemble of concentric, independently
evolving rings, progressively built up by infall of primordial composition. 
The assumption of infall is traditionally based upon the need to explain the
locally observed metallicity distribution of long-lived stars, which cannot
be explained by the simple ``closed-box'' model (leading to the well-known 
``G-dwarf problem''). However, the recent work of Blitz et al. (1999) gives 
observational support to this idea, showing that the Milky Way and M31 are 
currently accreting substantial amounts of gas ($\sim$ 1 \ms/yr) in the form 
of high velocity clouds of low metallicity.

The infall rate is assumed to be exponentially decreasing in time, i.e.
\begin{equation}
f(t,R) \  = \ A(R) \ e^{-t/\tau(R)}
\end{equation}
with a characteristic timescale $\tau(R_0$) = 7 Gyr in the solar
neighborhood ($R_0$ = 8kpc), in order to reproduce the local G-dwarf
metallicity distribution. $\tau(R)$ is assumed to increase outwards, 
from $\tau$(R = 2 kpc) = 1 Gyr to $\tau$(R = 17 kpc) = 12 Gyr. This radial 
dependence of the timescale of the infall rate
$f(R)$ is simulating the inside-out formation of galactic disks and, 
combined with the adopted SFR $\Psi(R)$ (Eq. 3)
allows to reproduce the observed current profiles
of gas, oxygen abundance and SFR in the Milky Way disk 
(see BP99 and Sects. 2.3 and 4 below). The coefficient
$A(R)$ is obtained by the requirement that at time T = 13.5 Gyr the current
mass profile of the disk $\Sigma(R)$ is obtained, i.e.
\begin{equation}
\int_0^T f(t,R) \  = \ \Sigma(R)
\end{equation}
with $\Sigma(R) \ \propto e^{-R/R_G}$ and a scalelength $R_G$ = 2.6 kpc for the
Milky Way disk.

The chemical evolution of each zone is followed by solving the appropriate
set of integro-differential equations, without the Instantaneous Recycling
Approximation. The adopted stellar Initial Mass Function (IMF)
is a multi-slope power-law between 0.1 \ms \ and 100 \ms \ from the work of
Kroupa et al. (1993), leading to a Return Fraction R = 0.32.

The star formation rate (SFR) is locally given by a
Schmidt-type law, i.e. proportional to some power of the gas surface density 
$\Sigma_g$: $\Psi \propto \Sigma_g^{1.5}$, according to the
observations of Kennicutt (1998). It varies with galactocentric radius $R$, as:
\begin{equation}
 \Psi(t,R) \ = \  \alpha \  \Sigma_g(t,R)^{1.5} \ V(R) \ R^{-1}
\end{equation}
where $V(R)$ is the circular velocity at radius $R$. This radial dependence of
the SFR is suggested by
the theory of star formation induced by density waves in spiral
galaxies (e.g. Wyse and Silk 1989). Since $V(R)\sim constant$ in the largest part 
of the disk, this is equivalent to $\Psi(R) \propto \Sigma_g(R)^{1.5} R^{-1}$.
The efficiency  $\alpha$ of the SFR in Eq. 3 is fixed by the requirement 
that the local gas fraction $\sigma_g(R_0)\sim$ 0.2, is reproduced at T = 13.5 Gyr.

We assume that the ``rings'' of the disk are evolving  independently from one
another. This (over)simplification
ignores in general the possibility of radial inflows in gaseous disks, 
resulting e.g. by viscosity  or by
infalling gas with specific angular momentum different from the one
of the underlying disk; in both cases, the resulting redistribution of 
angular momentum leads to radial mass flows. The magnitude of the effect is
difficult to evaluate, because of our poor understanding of viscosity 
and our ignorance of the kinematics of the infalling gas.
Models with radial inflows have been explored in the past
(Mayor and Vigroux 1981; Lacey and Fall 1985; Clarke 1989; 
Chamcham and Tayler 1994). It turns out that for some combinations
of the parameters of infall, radial inflow and SFR, acceptable solutions
are obtained, i.e. the current radial profiles of various quantities are
successfully reproduced (see, e.g. Portinari and Chiosi 2000 for 
a recent overview of the problem). However, at the present
stage of our knowledge introduction of radial inflows in the models
would imply more free parameters than observables. 
For simplicity reasons we stick to the model of ``independently evolving
rings'' for the Milky Way disk.

\subsection {Yields of massive stars}

An important ingredient in our study of abundance gradients are the stellar
yields of various elements. Most of the intermediate mass elements studied
here are produced by massive stars, with the exception of some CNO isotopes
that are also produced by intermediate mass stars. 
{\it We consider no yields from intermediate mass stars
in this work}; in the line of Goswami and Prantzos (2000), concerning the evolution of
the halo+local disk, our explicit purpose is to {\it check to what extent
massive stars can account for observations of intermediate mass elements
and for which elements the contribution of intermediate mass stars 
is mandatory.}

We use the metallicity dependant yields of WW95, which are given for stars 
of mass M = 12, 13, 15, 18, 20, 22, 25, 30, 35 and 40 \ms \ and metallicities
Z/\zs = 0, 10$^{-4}$, 10$^{-2}$, 10$^{-1}$ and 1. In Fig. 1 we
present the WW95 yields, folded with the Kroupa et al. (1993) IMF.  They are 
presented as {\it overproduction factors}, i.e. 
the yields (ejected mass of a given element) are divided by the 
mass of that element initially present in the part of the star 
that is finally ejected: 
\begin{equation}
<F> \ = \ {{\int_{M1}^{M2} Y_i(M) \ \Phi(M) \ dM} \over {\int_{M1}^{M2} 
X_{\odot,i}(M-M_R) \ \Phi(M) \ dM}}
\end{equation}
where: $\Phi(M)$ is the IMF, $M1$ and $M2$ the lower and upper mass limits
of the stellar models (12 \ms \ and 40 \ms, respectively).
$Y_i(M)$ are the individual stellar yields and $M_R$ the mass of the stellar 
remnant. Adopting $X_{\odot,i}$ in Eq. (4)
creates a slight inconsistency with the definition of the overpoduction factor
given above, but it allows to visualize the effects of metallicity in the yields
of secondary and odd-Z elements. 

\begin{figure}
\psfig{file=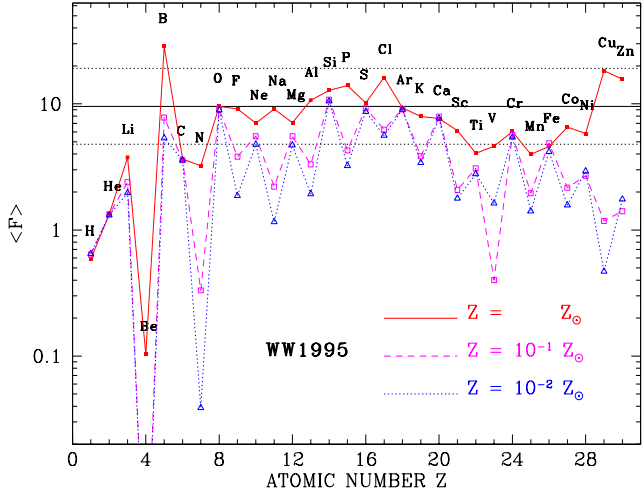,angle=-90,height=7.cm,width=0.5\textwidth}
\caption{\label{} 
Average overproduction factors (over a Kroupa et al. (1993) IMF, see Eq. 4) of 
the yields of Woosley and Weaver (1995) for 3 different initial stellar
metallicities, covering reasonably well the metallicity evolution of the
Milky Way disk. The {\it solid horizontal line} 
is placed at $F_{oxygen}$ and the two {\it dotted horizontal lines}
at half and twice that value, respectively.  Most of the intermediate mass 
elements are nicely co-produced (within a factor of two). 
The ``odd-even effect'' is clearly seen (e.g. in the cases of Na, Al, P, etc.); 
notice the small (but significant) metallicity dependence of the Ne and Mg yields. 
N behaves as a pure ``secondary'' element. The elements He, C, N, Li and Be 
obviously require another production site.  
}
\end{figure}

As can be seen from Fig. 1: i) most of the intermediate mass elements are nicely
co-produced (within a factor of 2) by solar metallicity
stars; ii) the ``odd-even effect'', favoring the production of odd-nuclei
at high metallicities, is clearly present; iii)
the yields of Ne and Mg show, curiously, some dependence on metallicity
(not as large as the one of the odd-elements Na and Al, but still enough to lead
to some interesting  abundance patterns, as we shall see in Sect. 3); iv)
He, C, N, Sc, V and Ti are underproduced relative to Oxygen. 
He, C and N clearly require another source (intermediate mass stars and/or Wolf-Rayet
stars, see Prantzos et al. 1994 and Sect. 4.2), while  the situation is less 
clear for the  elements, Sc, V and Ti (see Goswami and Prantzos 2000).

The calculations of WW95 did not consider any mass loss during stellar evolution.
Thus, they probably underestimated the yields of several elements that are 
expelled by the intense winds of massive stars, i.e. He, N and C. 
The effect of stellar winds is stronger when the stellar metallicity is larger.
Maeder (1992) found that stars with $M>$30 \ms \ and Z$>$0.1 \zs \ 
eject considerably larger amounts of He, N and C during their lifetime than their 
lower metallicity counterparts; for that reason, less matter is left in the He-core 
to be processed into Oxygen. At lower metallicities, the effect of stellar winds 
is negligible and stars of all masses evolve almost at constant mass. Elements
heavier than Oxygen are produced in the subsequent, very rapid, stages of
stellar evolution (after core He exhaustion) and their yields are not directly 
affected by the intensity of the mass loss. However, the structure of the stellar
core may be affected by the loss of mass and this may also affect the final
yields of heavy elements (e.g. Woosley, Langer and Weaver 1993).

\begin{figure}
\psfig{file=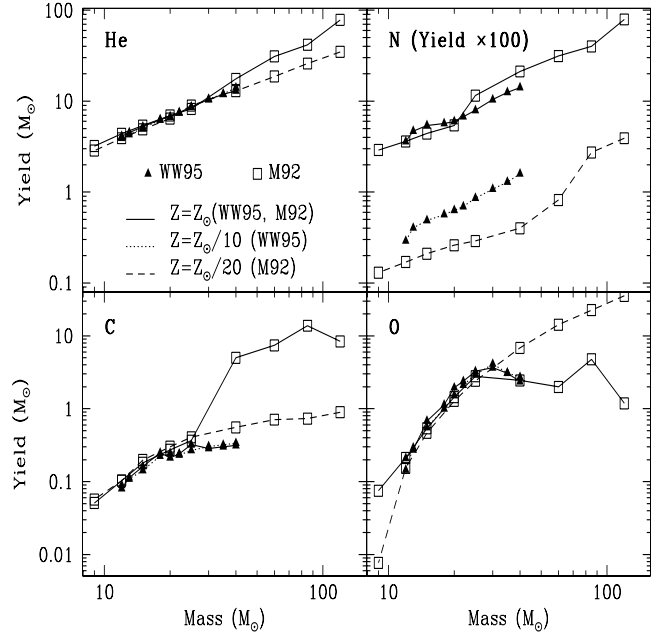,height=9.cm,width=0.5\textwidth,angle=-90}
\caption{\label{} 
Massive star yields of He, C, N and O for different initial
metallicities, according to WW95 ({\it filled triangles}) 
and M92 ({\it open squares}); {\it Solid curves}: Z = \zs, 
{\it dotted curves}: Z = 0.1 \zs \ (for WW95), {\it dashed curves:} 
Z = 0.05 \zs \ (for M92). There is good general agreement between the two 
calculations for stars with initial metallicity Z = \zs. He, C and O are 
primary elements, i.e. their yields are independant of 
the initial stellar metallicity; the differences between  the two sets 
of M92 yields for stars with  M$>$30 \ms \ are due to the metallicity 
dependent stellar winds. Nitrogen is produced as a secondary element
in both WW95 and M92.
}
\end{figure}

In Fig. 2 we present the yields of M92 for He, N, C and O as a function of stellar
mass; they are given for two metallicities (Z/\zs = 0.05 and 1, respectively) and
are compared to the corresponding yields of WW95. The  aforementionned effect
of metallicity-dependent stellar winds on the yields of stars with M$>$30 \ms \
is clearly seen. In Sect. 3.4 we shall explore the effect of those yields on the
abundance gradients in the disk.

To account for the additional source of Fe-peak elements,
required to explain the observed decline of O/Fe abundance ratio in the disk
(e.g. Goswami and Prantzos 2000), we utilise the recent yields of SNIa from 
the exploding Chandrashekhar-mass CO white dwarf models W7 and W70
of Iwamoto et al. (1999). These are updated versions of the original
W7 model of Thielemann et al. (1986), calculated for metallicities
Z = \zs (W7) and Z = 0 (W70), respectively. In this model, the deflagration 
is starting in the centre of an accreting white dwarf,  burns $\sim$ half 
of the stellar material in Nuclear Statistical Equilibrium
and produces $\sim$ 0.7 $M_{\odot}$ of $^{56}$Fe ( in the form of $^{56}$Ni).
These SNIa models lead to an oveproduction of Ni, but the evolution of
this element will no be considered here.

It should be emphasised that the evolution of the SNIa rate is not well
known, and hardly constrained by observations.
For the purpose of this work, we shall adopt the formalism of Matteucci 
and Greggio (1986) for the rate of SNIa, adjusting it as to have them 
appearing locally after the first Gyr, i.e. at a time when [Fe/H]$\sim$ $-$1 
in the solar neighborhood. However, use of that same formalism along the 
disk will lead to different O/Fe abundance ratios at T = 13.5 Gyr 
as we shall see in Sect. 3, i.e. the final Fe gradient will be different 
from the one of oxygen (see also Prantzos and Aubert 1995).

\subsection {Results for the Milky Way disk}

\begin{figure}
\psfig{file=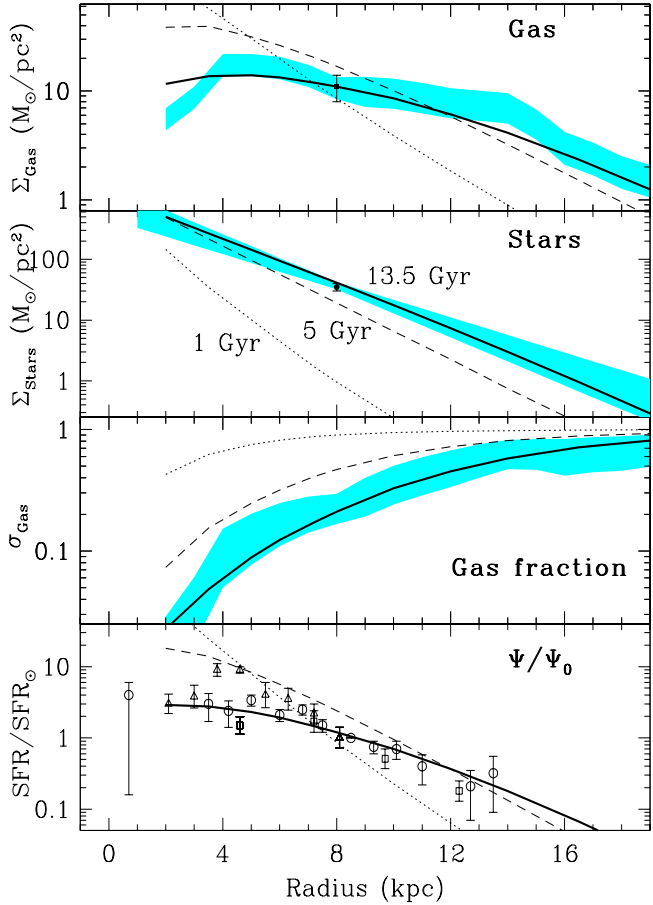,height=13.cm,width=0.5\textwidth}
\caption{\label{} 
Results of the chemical evolution model for the Milky Way disk and comparison 
to observations. In all panels, results are shown at three different epochs
({\it dotted curves:} 1 Gyr, {\it dashed curves:} 5 Gyr, 
{\it solid curves:} 13.5 Gyr). The latter are to be compared to observations 
of present-day profiles in the Milky Way disk, shown as {\it shaded regions} 
in first three panels and within {\it error bars} in the lower panel. 
In all panels, the error bar at $R_0$ = 8 kpc indicates observed quantities in the 
solar neighborhood. References for data are given in BP99.
}
\end{figure}

As described in detail in BP99, the simple model presented in Sect. 2.1 can readily
account for the evolution of the solar neighborhood, reproducing quite successfully
the main observational constraints (age-metallicity relationship, metallicity
distribution of long-lived F-stars, current local surface densities of stars, gas,
star formation and supernova rates). Also, in  Goswami and Prantzos (2000) it
is shown that the use of the WW95 yields for massive stars and of the 
Iwamoto et al. (1999) yields for SNIa leads to a successful agreement between the
gaseous composition of the model at an age of 9 Gyr and the observed solar one.
A few exceptions concern the elements He, C and N (which are underproduced) and Ni
(which is overproduced, because of the adopted SNIa yields).

The model predictions for the disk are equally successful, at least to a first order.
Indeed, the adopted combination of SFR (Eq. 3) and infall rate (Eq. 1) 
leads to final profiles of gas and SFR that are in fair agreeement 
with the observed ones, as can be seen in Fig. 3. The stellar profile is also
in agreement with observations, but it is essentially determined by the boundary
condition of Eq. 2. However, the adopted inside-out formation scheme of the disk
leads naturally to different scalelengths in the B-band
(reflecting mostly the SFR profile in the past $\sim$ 1 Gyr) and the K-band
(reflecting the total stellar population, cumulated over T = 13.5 Gyr). As 
shown in BP99, the corresponding scale-lengths ($\sim$ 4 kpc in the B-band and
$\sim$ 2.6 kpc in the K-band, respectively) are in fair agreement with
observations. Moreover, the model also reproduces reasonably well the total
current SFR and supernova rates  as well as the total luminosities in various 
wavelenght bands. This is a rather encouraging success, since the number of the 
constraints is much larger than the number of the parameters. The success of 
this simple  model encourages us to use it for a thorough study of the various 
abundance gradients in the Milky Way disk.

\section{Model results for abundance gradients}

We calculated the evolution of the abundances of all elements between H and Zn 
with the WW95 metallicity dependent yields in all the zones of our model disk.
In Fig. 4 we present the results concerning all elements with measured
abundance gradients in the Milky Way. We notice that, since we did not include
yields from IMS in our calculation, our results for He, C and N represent rather
lower limits. We shall discuss the comparison to observations in the next section, 
where we shall also explore the role of massive, mass losing stars to the abundance 
profiles of those elements. Here we focus on the model results, which may be
summarised as follows:

i) Final values (at T = 13.5 Gyr) of the abundances at R$_0$ = 8 kpc are $\sim$ solar
for O, Al, Si, S, Ar and Fe; they are slightly lower than solar for Ne and Mg; and
they are considerably below solar for C and N. It may appear surprising that
Ne and Mg do not reach their solar values, but this is a consequence of the adopted
metallicity dependent yields of WW95 (Fig. 1): the {\it average} (i.e. over the 
disk's age) overproduction factors of Ne and Mg are lower than the one of e.g. O.
It is not clear whether such a dependence is physical or just an artifact of the
WW95 models. As for C and N, it is clear that massive stars with no mass loss
cannot be the main sources of those elements (see also Sect. 4).

\begin{figure}
\psfig{file=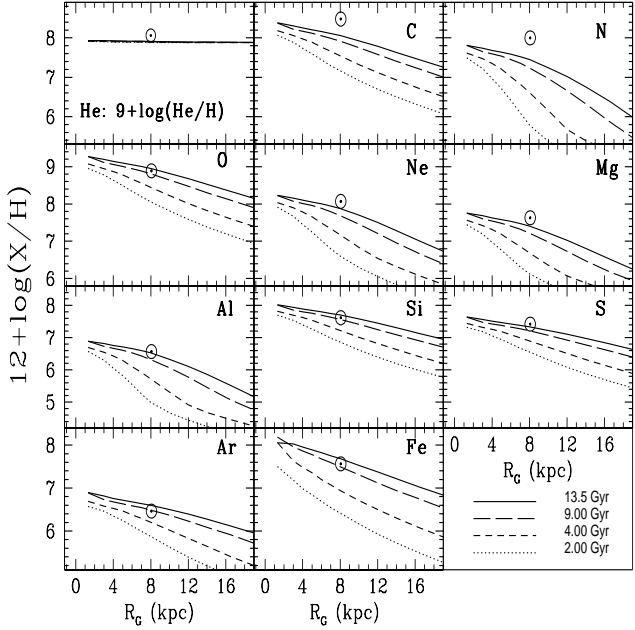,angle=-90,height=9.cm,width=0.5\textwidth}
\caption{\label{} 
Evolution of abundance profiles in the Milky Way disk, in the framework
of the model presented in Sect. 2 and with the WW95 metallicity dependent yields.
Curves show results at 2, 4, 9 and 13.5 Gyr, respectively (as indicated in the
lower right panel). Because of the metallicity dependence of the yields,
some elements (N, Al) have steeper abundance profiles than e.g. O, at all times.
This is also the case for Fe, but the reason is its production mostly by SNIa.
C and N are underproduced with respect to their solar system values, because 
yields from intermediate mass stars are neglected in this calculation. Metallicity
dependent massive star yields from M92 considerably improve the situation
concerning carbon (see Fig. 7), but not nitrogen. Solar abundances ($\odot$) 
were taken from Grevesse \& Sauval (1998).
}
\end{figure}

ii) The most prominent feature of the model is the prediction that abundance
gradients flatten with time. This is a generic feature of all models
forming the galactic disk ``inside-out''. Indeed, in that case, there is a rapid
increase of the metal abundance at early times in the inner disk, leading to a
steep abundance gradient. As time goes on, star formation ``migrates''
to the outer disk, producing metals there and flattening the abundance gradient.

iii) The final abundance profile (T = 13.5 Gyr) is, in general, flatter in the 
inner disk. As already described in Prantzos and Aubert (1995) this is due to 
the fact that in those regions the large populations of low-mass, long-lived 
stars that are formed early on in galactic history reject a lot of metal-poor 
gas at the end of their evolution, which dilutes the metal abundances; this 
effect is absent in the outer regions, where there are not very old stellar 
populations. Notice that the flattening seems to be more important in the case 
of N, Ne, Mg and Al. These elements show some metallicity dependence in their 
yields (at least according to WW95) which is difficult to understand in the 
case of Ne and Mg, but expected in the case of the secondary N and of the odd-Z 
Al. Since WW95 give yields only up to stellar metallicities of Z = \zs, for 
higher metallicities we use their Z = \zs yields. This means that in the inner 
disk, where metallicities higher than solar are reached, we underestimate the 
production of any element with metallicity dependent yield. If appropriate yields 
were used, the  Al and N profiles in the inner disk would be steeper, not flatter, 
than the one of O.

\begin{figure}
\psfig{file=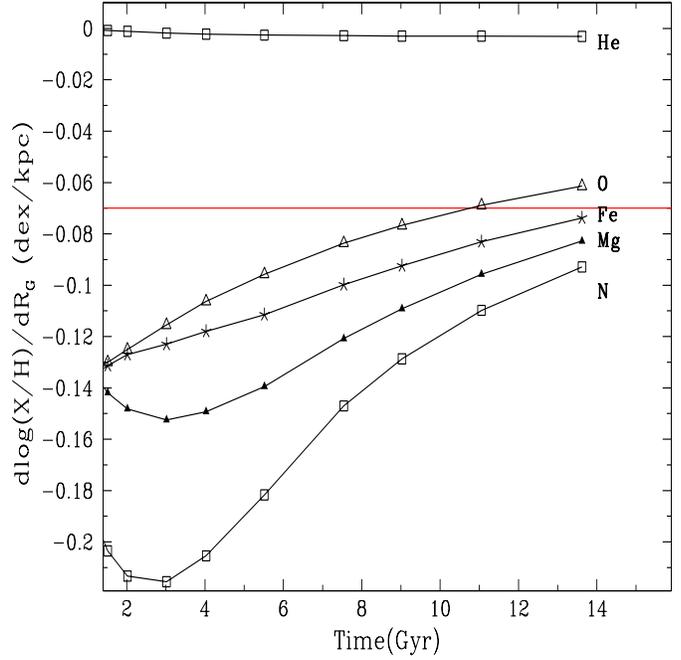,angle=-90,height=9.cm,width=0.5\textwidth}
\caption{\label{} 
Evolution of abundance gradients ( expressed in dex/kpc )
in the Milky Way disk (range: 4$-$14 kpc from the centre), in the framework
of the model presented in Sect. 2 and with the WW95 metallicity dependent 
yields. Because of the inside-out formation scheme adopted here, gradients 
were steeper in the past. Because of the metallicity dependence of the 
yields of WW95, some elements, like N or Mg, have larger abundance gradients 
(in absolute value) than O. This is also the case for Fe, but the reason 
is its production mostly by SNIa.  }
\end{figure}

The magnitude of the current abundance gradients in the Milky Way disk is one
of the most important constraints in the models of the evolution of our Galaxy.
Most of the proposed models reproduce it fairly well (e.g. Tosi 2000 and references
therein), at least when no radial inflows are included. However, equally
important is the question of the evolution of those gradients and, in particular,
whether they flatten or steepen with time. We shall confront our models to
the data and to other theoretical works in the next section. Here, we present
the evolution of our model gradients in the 4$-$14 kpc region for a few selected
elements (Fig. 5). All gradients were systematically larger in the past. 
The gradient of Fe is slightly larger than the one of O, because our adopted 
prescription for the SNIa rate produces a smaller O/Fe ratio in the inner disk 
than in the outer one. The gradient of secondary N is always steeper than the 
one of O, but since we do not include N production from intermediate mass stars 
or WR stars in this calculation, this result serves merely for illustration purposes. 

\section{Comparison to observations} 

In this section we compare our results to the observed abundance profiles of 
various elements across the galactic disk. Tracers of abundance profiles include
emission-line objects (HII regions and planetary nebulae), as well as stars and
stellar associations (B-stars and open clusters, respectively).
In most cases, those tracers are relatively young objects (HII regions, B-stars
and planetary nebulae of type I), younger than $\sim$ 1 Gyr; they provide then 
information about the current abundance profile of the Milky Way disk.
In other cases (planetary nebulae of type II or III, open clusters),
the objects involved are several Gyr old and provide information about the past
status of the disk (albeit with much larger uncertainties than in the former case).

In Sect. 4.1 we present briefly the available observational data from various sources.
In Sect. 4.2 we compare the data for young objects  to our results at T = 13.5 Gyr; 
we show, in particular, how the results for C (and to a much lesser extent, N and O) 
may be affected by the M92 metallicity-dependent yields of massive, mass losing stars. 
In Sect. 4.3 we discuss the profiles of the corresponding abundance ratios of elements 
to Oxygen.
In Sect. 4.4 we discuss the history of the abundance profiles, comparing our
results to  observations of old objects (of rather uncertain ages).

\subsection{Observational data}

One of the first comprehensive optical surveys of HII region abundances was performed 
by Shaver et al. (1983). They have found strong abundance gradients for N/H, O/H and 
Ar/H in the range of galactocentric distances R$_G$ = 5 $-$ 12  kpc. Those results 
were later confirmed by the surveys of Fich \& Silkey(1991) and V{\'\i}lchez \& 
Esteban (1996), 
extending the data up to distances of R$_G\sim$ 17 kpc. Those authors suggested 
that the N/H and O/H gradients show a tendency for flatenning between 11 and 18 kpc.
Taking advantage of the small extinction in the infrared, Simpson et al. (1995) and 
Afflerbach et al. (1997) studied objects towards the center of the Galaxy. Simpson 
et al. (1995) observed 12 HII regions in R$_G$ = 0 $-$ 10 kpc and found a somewhat 
better fit with a step function than with a smoothly decreasing one. On the contrary, 
after adding five outer Galaxy HII regions (R$_G$ = 13 $-$ 17 kpc), Rudolph et al. 
(1997) concluded that the single slope profiles for N/H and S/H are more likely than 
step functions. The existence of a gradient in the oxygen abundance profile obtained 
from HII regions is not in doubt at present, but its magnitude has been recently 
challenged by the work of Deharveng et al. (2000), who report a value about half 
as large as previous measurements. 
 
Observations of  B-type stars in young clusters and associations in the late 90ies 
confirmed the existence of an abundance gradient. Smartt \& Rolleston (1997) found 
an oxygen abundance gradient similar to the one obtained in HII regions (except for 
the work of Deharveng et al. 2000), in the galactocentric distance range 
R$_G$ = 6 $-$ 18 kpc. This was confirmed by Gummersbach et al. (1998) who detected 
significant abundance gradients for C, O, Mg, Al  and Si in galactocentric distances 
R$_G$ = 5 $-$ 14 kpc.

\begin{table*}[h]

{\footnotesize
\noindent
{\bf TABLE 1.  Observed and calculated abundance gradients in the Galactic disk (dlog(X/H)/dR, in dex/kpc)} \\ [2mm]
\begin{tabular}{lccccccccccccc} 
\hline \hline
Object   &  Rg   &    He      &   C  &    N       &   O        & Ne & Mg & Al & Si &  S        &   Ar       & Ref. \\ 
class    & (kpc) &            &      &            &            &    &    &    &    &           &            &      \\ 
\hline
 HII & 4--14 & $-$0.001     &      &   $-$0.090   &  $-$0.070    &            & & & &   $-$0.010 & $-$0.060   & (1)  \\
     &       & $\pm$0.008   &      & $\pm$0.015   & $\pm$0.015   &            & & & & $\pm$0.020 & $\pm$0.015 &      \\
     &   &            &      &            &            &    &    &    &    &           &            &      \\ 
 HII & 0--10 &            &      &   $-$0.100   &            & $-$0.080     & & & &   $-$0.070   &            & (2)  \\ 
     &       &            &      & $\pm$0.020 &            & $\pm$0.020 & & & & $\pm$0.020 &            &      \\
     &   &            &      &            &            &    &    &    &    &           &            &      \\ 
 HII & 0--17 &            &      &   $-$0.111   &            &            & & & &   $-$0.079   &            & (3)  \\
     &       &            &      & $\pm$0.012 &            &            & & & & $\pm$0.009 &            &      \\
     &   &            &      &            &            &    &    &    &    &           &            &      \\ 
 HII & 0--12 &            &      &   $-$0.072   &   $-$0.064   &            & & & &   $-$0.063   &            & (4)  \\
     &       &            &      & $\pm$0.006 & $\pm$0.009 &            & & & & $\pm$0.006 &            &      \\
     &   &            &      &            &            &    &    &    &    &           &            &      \\ 
 HII & 5--15 &            &      &            &   $-$0.040   &            & & & &            &            & (5)  \\
     &       &            &      &            & $\pm$0.005 &            & & & &            &            &      \\
     &   &            &      &            &            &    &    &    &    &           &            &      \\ 
 HII & 6--9  & $-$0.004     &   $-$0.133   &   $-$0.048   &  $-$0.049    &   $-$0.045   & & & &  $-$0.055    &   $-$0.044   & (6) \\ 
     &       & $\pm$0.005 & $\pm$0.022 & $\pm$0.017 & $\pm$0.017 & $\pm$0.017 & & & & $\pm$0.017 & $\pm$0.030 &   \\
\hline
 B stars & 6--18 &          &      &            &   $-$0.070   &            & & & &            &            &   (7)  \\ 
         &       &          &      &            & $\pm$0.010 &            & & & &            &            &      \\
     &   &            &      &            &            &    &    &    &    &           &            &      \\ 
 B stars & 5--14 & &   $-$0.035   &  $-$0.078    &  $-$0.067    & &  $-$0.082    &   $-$0.045   &  $-$0.107    &   &  & (8)\\ 
         &       & & $\pm$0.014 & $\pm$0.023 & $\pm$0.024 & & $\pm$0.026 & $\pm$0.023 & $\pm$0.028 &   &  &      \\
\hline 
 PNI   & 4--14 &          &      &            &   $-$0.030   &   $-$0.004   & &  &  &   $-$0.075   &   $-$0.060   & (9) \\
       &       &          &      &            & $\pm$0.007 & $\pm$0.008 & &  &  & $\pm$0.008 & $\pm$0.008 &    \\
     &   &            &      &            &            &    &    &    &    &           &            &      \\ 
 PNII  & 4--14 &          &      &            &   $-$0.058   &   $-$0.036   & &  &  &   $-$0.077   &   $-$0.051   & (10) \\
       &       &          &      &            & $\pm$0.007 & $\pm$0.010 & &  &  & $\pm$0.011 & $\pm$0.010 &    \\
     &   &            &      &            &            &    &    &    &    &           &            &      \\ 
 PNIII & 4--14 &          &      &            &   $-$0.058   &   $-$0.041   & &  &  &   $-$0.063   &   $-$0.034   & (9) \\
       &       &          &      &            & $\pm$0.008 & $\pm$0.008 & &  &  & $\pm$0.010 & $\pm$0.010 &     \\
\hline
 All &   & &   $-$0.07   &  $-$0.08    &  $-$0.07    & &  $-$0.07    &   $-$0.05   &  $-$0.06    & $-$0.07  &  & (11) \\ 
         &       & & $\pm$0.02 & $\pm$0.02 & $\pm$0.01 & & $\pm$0.01 & $\pm$0.02 & $\pm$0.01 & $\pm$0.02  &  &      \\
\hline \hline
Model$^a$ & 4--14 &  $-$0.003  & $-$0.062 & $-$0.093 & $-$0.061 & $-$0.082 & $-$0.083 & $-$0.090 & $-$0.058 & $-$0.053 & $-$0.048 &    \\
Model$^b$ & 4--14 &  $-$0.004  & $-$0.086 & $-$0.096 & $-$0.058 &        &        &        &        &        &        &    \\
\hline \hline

\end{tabular} \\[2mm]

\noindent
{\bf References:}
(1) Shaver et al. 1983; 
(2) Simpson et al. 1995; 
(3) Rudolph et al. 1996; 
(4) Afflerbach et al. 1997; 
(5) Deharveng et al. 2000; \\
(6) Esteban et al. 1999; 
(7) Smartt \& Rolleston 1997; 
(8) Gummersbach et al. 1998; 
(9) Maciel \& Koppen 1994; 
(10) Maciel \& Quireza 1999;
(11) Smarrt (2000) \\
\noindent
{\bf Results}:
$^a$ Woosley and Weaver (1995) yields;
$^b$ Maeder (1992) yields for He, C, N, O \\ 

}
\end{table*}

Several works have been devoted to the study of the abundance patterns of planetary 
nebulae (PN, see e.g. Maciel and Quiroza 1999 and references therein). 
According to a classification scheme originally suggested in Peimbert (1978) 
and revised in Pasquali and Perinotto (1993), planetary nebulae are subdivided 
into types I, II and III. PNI have He/H $>$ 0.125 and log(N/O) $>$ $-$0.3, 
are located in the galactic thin disk and their progenitors are probably 
associated with stars in the mass range 2.5 $-$ 8 \ms. PNII belong also to the 
thin disk, are not particularly enriched in nitrogen or He and are thought 
to evolve from stars in the 1.2 $-$ 2.5 \ms \ range. Finally, PNIII are associated 
with the galactic thick disk and thought to originate from low mass progenitors, 
in the 1 $-$ 1.2 \ms \ range. The estimated progenitor masses imply ages of $<$1 Gyr 
for PNI, 1 $-$ 8 Gyr for PNII and $>$8 Gyr for PNIII, respectively. 
These differences should allow, in principle, to use PN as tracers of the 
chemical evolution of the galactic disk. However, the mass and age differences
between the various PN types are not quite well defined, as we shall see in 
Sect. 4.3. Moreover, all PN are expected to be auto-enriched in products of 
the CN cycle (i.e. N-rich and C-poor), while PNI (especially those originating 
from relatively massive progenitors, in the 5 $-$ 8 \ms \ range) are probably 
auto-enriched in products of the NO cycle (N-rich and O-poor). For those reasons, 
we shall not consider at all in the following the abundances of C and N from PN 
of all types and  the abundances of O from PNI. For the purpose of this work, 
we shall use PNI (along with HII regions and B-stars) as tracers of the young 
disk in Sects. 4.2 and 4.3. PNII and PNIII will be discussed in Sect. 4.4.

Table 1 summarizes the currently available observational data on the abundance
profiles across the Milky Way disk. Two points should be noticed:

1) Contrary to other authors, we did not
consider data on Fe from open clusters, since the situation is rather uncertain
at  present: for instance, in the solar neighborhood, open cluster data show
no evolution of the Fe abundance with age (e.g. Friel 1999), 
contrary to the familiar age-metallicity relation suggested by data for F-stars
(Edvardsson et al. 1993). In view of that dicrepancy, we chose not to consider
Fe in this work (although we do calculate its abundance profile, as shown in 
Fig. 4).

2) Data in Table 1 are adopted  directly from the references 
listed and no attempt has been made to homogenize them by recalculating the abundances 
in a consistent way. In general, differences in techniques and atomic data employed 
produce an abundance scatter which is smaller than observational uncertainties in the 
line strengths. Therefore, we believe that direct comparison between our models
and those inhomogeneous  data  can still provide statistically meaningful results.

\subsection {Current disk abundance profiles}

\begin{figure*}
\psfig{file=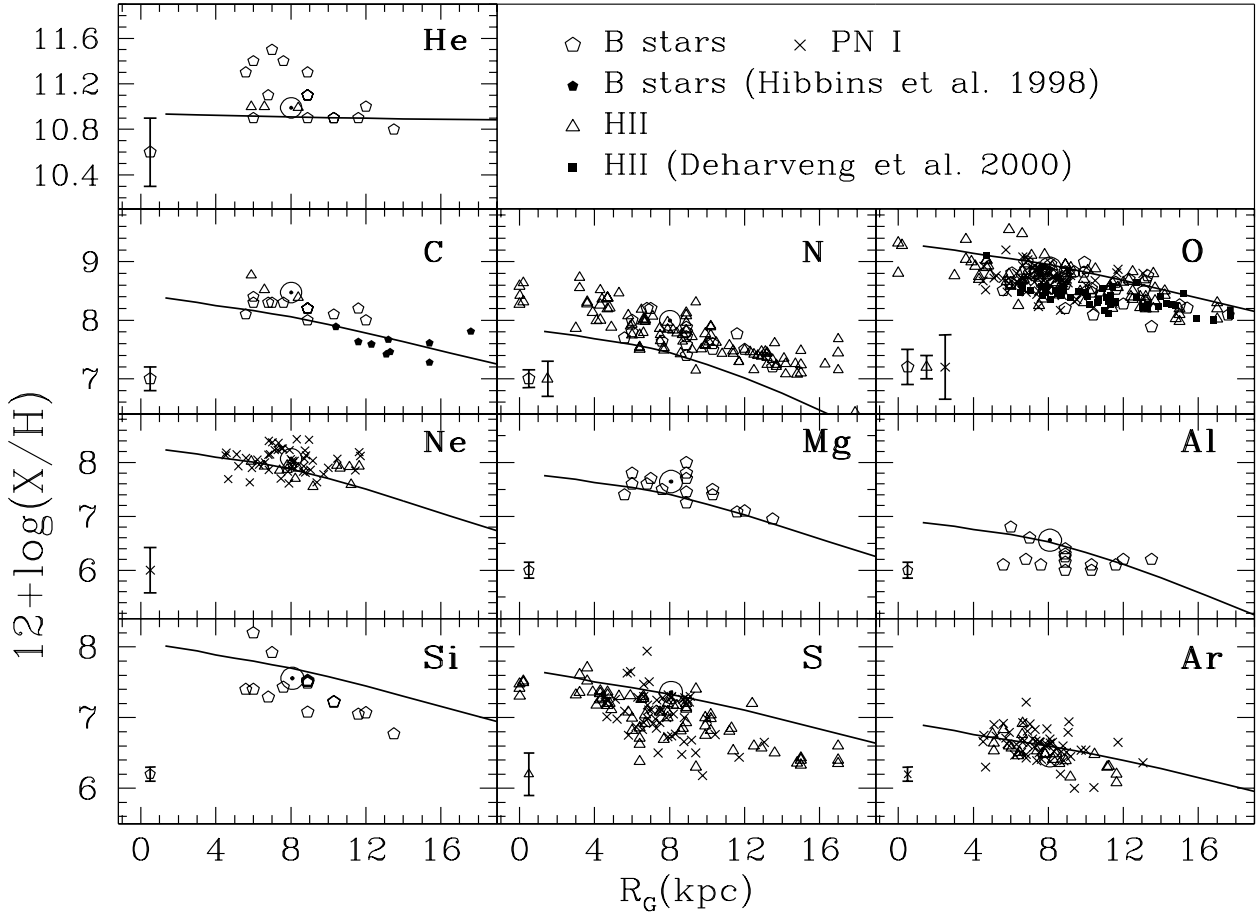,angle=-90,height=13.cm,width=\textwidth}
\caption{\label{} 
Present day abundance profiles in the Milky Way disk (obtained in the framework
of the model presented in Sect. 2 and with the WW95 metallicity dependent yields)
and comparison with observations.
Data sources are given in Table 1. The HII region data of Deharveng et al. (2000)
suggest an oxygen abundance gradient by 40\% 
less steep than the commonly accepted one of dlog(O/H)/dR $\sim$ $-$0.065 dex/kpc.
The typical observational uncertainties for different elements and data sets are 
shown with error bars in the bottom left corner of each panel.
}
\end{figure*}

The observational data of Table 1 concerning ``young'' objects (HII regions,
B-stars, PNI) are plotted in Fig. 6, along with our model results, obtained with
the WW95 yields, at T = 13.5 Gyr. In Fig. 7 we also plot the same data for He, C, N
and O and we show the corresponding results obtained with both WW95 and M92
yields; as discussed in Sect. 2.2, the metallicity dependenence of massive star winds 
has an effect on the yields of those elements. 

Since the prescriptions for the radial dependence for infall and  SFR of the 
BP99 model adopted here were such as to reproduce the observed oxygen 
abundance profile, we start the description of our results with this element.

{\sc Oxygen}: As can be seen from Table 1, data from both HII regions and
B-stars suggest an abundance gradient dlog(O/H)/dR $\sim$ $-$0.07 \dkp \ (see
the recent review by Smarrt 2000 on B-stars). Maciel and Quiroza (1999) 
include  PNII (despite the fact that these objects are, in principle, older 
than 1 Gyr) and find an average gradient of dlog(O/H)/dR $\sim$ $-$0.065 \dkp 
for all tracers of the ``young'' population. 
However, Deharveng et al. (2000), after a consistent analysis of their own 
data on HII regions, as well as of those of previous works, conclude
that the gradient should be $\sim$ 40\% 
smaller than generally thought, i.e. $-$0.04 \dkp. 
The data of Deharveng et al. (2000) concern the galactocentric distance 
R$_G$ = 5 $-$ 15 kpc, i.e. the same as the one of the work of Shaver et al. (1983) 
on HII regions or studies on B-stars; the difference in the resulting abundance 
gradient cannot then be attributed in the studied galactocentric distance range. 
As stressed by Deharveng et al. (2000), their derived O/H abundances depend 
strongly on their two-temperature HII region model (one temperature for the 
high excitation O$^{++}$ zone and another temperature for the low excitation 
O$^+$ zone); an alternative HII region model (with temperature fluctuations) 
would lead to larger O/H abundances.

As explained in Sect. 2, our model is based on the radial variation of the SFR 
(Eq. 3) and infall rate (Eq. 1). A different prescription would lead to different
results for the abundance gradients (as in e.g. Prantzos and Aubert 1995, who 
used a radially independent infall time-scale and found a smaller gradient of 
dlog(O/H)/dR $\sim$ $-$0.03 \dkp). Also, by adopting the same prescriptions, but
assuming radial inflows, one would obtain smaller values for the abundance gradient
(e.g. Tsujimoto et al. 1995, Portinari and Chiosi 2000). Obviously, the magnitude
of the abundance gradient is crucial in fixing the parameters entering the
current phenomenological chemical evolution models. In particular, a strong 
gradient does not support the idea of important radial inflows induced by a 
galactic bar (such as those found e.g. in the calculations of Friedli and Benz 1995); 
the Milky Way bar must then have formed relatively recently and/or played a
negligible role in driving gaseous flows in the disk.

The only published successful chemodynamical model for the Milky Way (Samland
et al. 1997) predicts an oxygen abundance gradient that flattens considerably 
in the outer disk. Despite some claims for  such a flatenning (Vilchez and 
Esteban 1996), no convincing observational evidence exists up to now (see the 
discussion in Deharveng et al. 2000). Our model predicts no flatenning of the 
gradient in the outer disk. It does predict a small flatenning in the inner disk. 
As explained in Prantzos and Aubert (1995), this is due to the fact  that all 
metal abundances are diluted in the inner disk by the late ejection of metal-free
material by the numerous low-mass, long-lived stars of the first stellar
generations; the ejection rate of that material at late times is larger than
the metal production rate, since most of the gas in the inner disk has been
consumed at that time. This is not the case in the outer disk, which is formed
late.

As can be seen in Fig. 7, the M92 yields lead to results for oxygen that are
almost indistinguishable from those of WW95. Only in the inner (more metal-rich) 
zones a slight difference is obtained in the oxygen abundance profile. As 
already discussed in Prantzos et al. (1994) the metallicity dependent stellar 
winds have a negligible impact on the oxygen yields.

Finally, another aspect of the observed oxygen abundances (Fig. 6) has been 
discussed in several places:  all ``young'' objects in the local disk have 
lower oxygen abundances (in the range O/H $\sim$ 3.2$-$5 10$^{-4}$) than the 
Sun ((O/H)$_{\odot}$ = 6.8$^{+1.0}_{-0.9}$ 10$^{-4}$, Grevesse \& Sauval 1998).
For instance, Gies and Lambert (1992) derive an oxygen abundace O/H$\sim$ 
4.8 10$^{-4}$ for local B stars, while Esteban et al.(1998) give O/H  
= 5.2$^{+0.9}_{-0.7}$ 10$^{-4}$ for Orion nebula.
This is difficult to understand in the framework of conventional models 
of chemical evolution, where metallicity increases monotonically with time. 
The idea that the Sun was born in the  metal-rich inner disk and subsequently 
migrated outwards  (Wielen et al. 1996), has been recently rejected by 
Binney and Sellwood (2000), on the grounds of dynamical arguments. Thus, 
at present, there is no satisfactory explanation for the ``super-metallicity'' 
of the Sun relative to young objects in the solar neighborhood.

{\sc Carbon}: There are very few studies of carbon abundances, either 
in HII regions or B-stars. Based on only two nearby HII regions 
(Orion and M17), Peimbert et al.(1992) first derived an abundance 
gradient of dlog(C/H)/dR $\sim$ $-$0.08$\pm$0.02 \dkp in the solar vicinity. 
Adding another HII region (M8) to that sample, Esteban et al. (1999) obtain  
$-$0.133$\pm$0.022 \dkp. However, both those studies concern a very limited 
range of galactocentric distances (6 $-$ 9 kpc), and the derived abundance 
gradients cannot be considered as representative of the disk as a whole. 
On the other hand, the study of Gummersbach et al. (1998) suggests a 
rather small C gradient for B-stars in the range R$_G$ = 6 $-$ 12 kpc:  
dlog(C/H)/dR $\sim$ $-$0.045 \dkp; this is half as steep as the oxygen 
abundance gradient derived for those same B-stars in that study. 
If the Gummersbach et al. (1998) value is correct, it should then be 
difficult to understand the corresponding disk profile of the C/O ratio, 
which should decrease in the inner disk. Indeed, observations of low 
mass stars in the solar neighborhood show that C/O increases with metallicity 
(e.g. Gustafsson et al. 1999 and references therein), and we would expect 
to see the same effect in the inner, metal-rich disk. Hibbins et al. (1998) 
performed a differential analysis of C and N abundances of B-stars in the 
outer disk (10 $-$ 17 kpc). They found that C correlates with N, 
but not with O. However, their absolute values of C/H for nearby stars are 
systematically lower by a factor 2 $\sim$ 3 than those of other studies; 
they suggest then that their data (and data coming from different samples, 
in general) should be used with caution in studies of galactic chemical 
evolution.

\begin{figure}
\vspace{-0.92cm}
\psfig{file=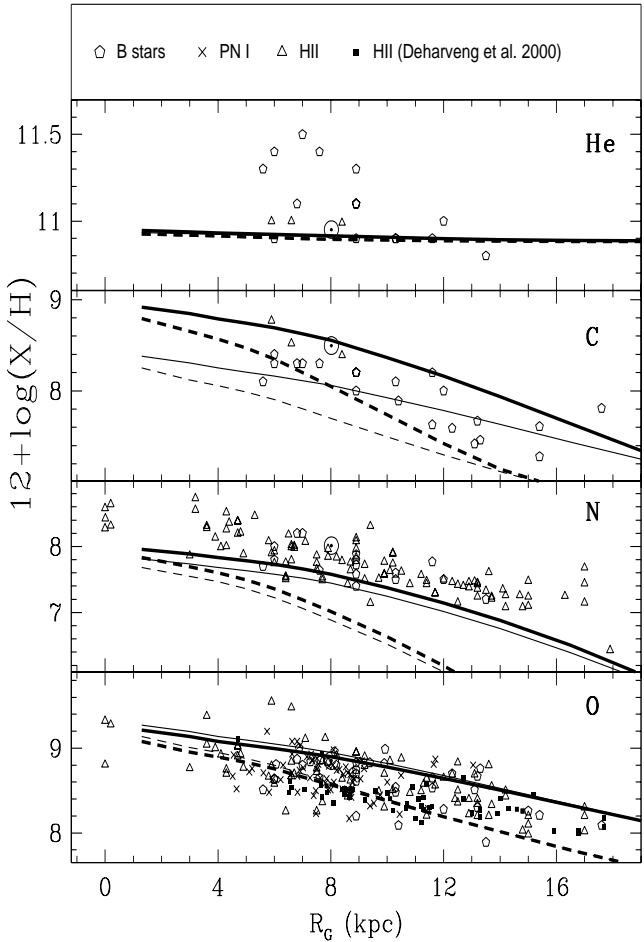,height=15.cm,width=0.5\textwidth}
\caption{\label{} 
Abundance profiles of He, C, N and O, at T = 9 Gyr ({\it dashed curves}) 
and T = 13.5 Gyr ({\it solid curves}), obtained with the WW95 yields 
({\it thin curves}, same as e.g. in Fig. 3), and the M92 yields 
({\it thick curves}), in the framework of the model presented in 
Sect. 2. It can be seen that, compared to the WW95 yields, the M92 yields 
lead to considerably different results for C, to slightly different results 
for N, and to quasi-identical results for He and O. In particular, the M92 
yields may explain completely the carbon abundance profile, with no need 
for any contribution from intermediate mass stars; on the contrary, nitrogen 
observations suggest that another source is mandatory.
}
\end{figure}

In summary, the carbon abundance profile in the Milky Way disk is poorly determined.
The few available data sets cover limited galactocentric distance ranges and are
not derived in a consistent way. If all the data are plotted, as in Fig. 6 and 7,
a carbon abundance gradient larger than the one of oxygen appears; but this may
well be an artifact. We note that in his recent review, Smartt (2000) suggests a 
gradient of  dlog(C/H)/dR $\sim$ $-$0.07 dex kpc$^{-1}$, 
based on unpublished work of Rolleston et al. (2000) on B-stars.

Our model results (Fig. 6) show a carbon abundance gradient of 
dlog(C/H)/dR $\sim$ $-$0.06 \dkp, i.e. comparable to the one of oxygen, 
when the WW95 yields are used. However, 
the absolute value of the local carbon abundance at the Sun's birth is underproduced 
in that case (by a factor of $\sim$2); another C source is then required. Intermediate 
mass stars (IMS) are well known net producers of carbon, but in recent years, several 
works (Prantzos et al. 1994, Carigi 1994, Gustaffson et al. 1999, Henry et al. 2000) 
suggested that the M92 metallicity dependent yields of massive stars fit better the 
observed C/O abundance patterns in extragalactic HII regions and low-mass stars in 
the solar neighborhood. This is confirmed in Fig. 7, where it can be seen that the 
use of the M92 yields allows indeed to reproduce the absolute abundance of carbon
in the solar neighborhood, with no need for a contribution by IMS. In that case, 
the resulting C abundance gradient is much steeper: $-$0.086 \dkp. Clearly, the 
abundance profiles of carbon vs. oxygen in the Milky Way disk are crucial in any 
attempt to evaluate the role of massive star winds in the production of carbon.

{\sc Nitrogen:} There is a large number of works concerning N abundances 
in HII regions; the N abundance profile seems to be steeper, in general, 
than the corresponding O profile (see Table 1). Observations of B-stars 
(Gummersbach et al. 1998) support this conclusion, but the error bars are 
much larger in that case. Smarrt (2000) suggests a gradient of  
dlog(N/H)/dR $\sim$ $-$0.08$\pm$0.002, i.e. compatible with the one of oxygen.
Our models also produce a steeper gradient of N (with respect to that of O),
with the yields of both WW95 (Fig. 6) and M92 (Fig. 7). As can be seen in 
Fig. 7, the differences resulting from the use of those two sets of yields 
are rather small (slightly larger than in the case of oxygen, but considerably
smaller than in the case of carbon). Two important points should be noticed:

i) The absolute value of the N abundance obtained with both sets of yields is
too low compared with observations in the solar neighborhood and in the disk. Contrary
to the case of carbon, the M92 yields are not sufficient to account for the current
abundance of N in the Milky Way. Unless the N yields are seriously underestimated
in both WW95 and M92 (which is improbable, since they consist of the sum of the
initial C+N+O), another N source is required. Intermediate mass stars are the obvious
candidate, but their yields are notoriously difficult to estimate (in view of the
many uncertainties concerning mass loss rates, ``hot-bottom burning'', etc.).
Our calculations clearly show what is the magnitude of the expected contribution
of IMS, in order to complement the (presumably better understood) yields of
massive stars: on average, IMS have to produce 3-4 times more N than massive stars.

ii) We stress again (see also Sect. 3) that the flatenning of the N abundance 
profile obtained in our calculations for the inner disk is due to the fact that,
since yields of WW95 and M92 are available only up to Z = \zs, for higher
metallicities we use the yields at Z = \zs; thus, in our models
 N is produced in the inner disk (where high metallicities are developed)
with the same yield always, i.e. as
a primary element and its abundance follows the one of oxygen. Clearly, this is
an artifact of the calculation, due to the lack of appropriate input data; if N
were treated correctly, as a secondary,  its abundance profile would not
flatten in the inner disk [{\it Notice}: we could simply scale the N yields
of WW95 with metallicity, but it would be more difficult to do the same
thing with the M92 yields, since the effect of the stellar wind on the yield
is not simply proportional to metallicity].

{\sc Magnesium, Aluminium, Silicon}: The abundance profiles of those elements
have been studied by Gummesrbach et al. (1998), who observed 
B-stars in galactocentric distances 5 $-$ 14 kpc. In view of the large
uncertainties (see Table 1), all those gradients can be considered as compatible
with the oxygen gradient of $-$0.07 \dkp, at the 1 $\sigma$ level; 
this is particularly true for Mg. On the other hand, taken at face value, 
the Al profile seems to be considerably steeper than the one of 
Si ($-$0.045 \dkp \ vs. $-$0.107 \dkp, respectively).
This is rather surprising, taking into account that Al is an odd-Z element
and its yield depends slightly on metallicity (the ``odd-even'' effect, see Fig. 1).
Because of this effect, our models produce a steeper gradient for Al than for Si.
The latter is comparable to the one of oxygen, as expected.

We note that in his recent review Smarrt (2000) suggests gradients of 
similar magnitude for Al and Si ($-$0.05$\pm$0.02 \dkp \ 
and $-$0.06$\pm$0.01 \dkp, respectively), based on B-star data. 
In that case the problem is alleviated, although not completely solved.
We also note that low-mass stars  in the solar neighborhood do not exhibit
the theoretically expected behaviour of the Al/O abundance pattern (e.g. Goswami
and Prantzos 2000 and references therein). It may well be then that the odd-even
effect has been overestimated in the WW95 yields of Al.

In summary, the observed abundance profiles of Al and Si run opposite to
theoretical expectations, but in view of their large uncertainties, it is
difficult to draw firm conclusions on stellar nucleosynthesis.
 
\begin{figure}
\psfig{file=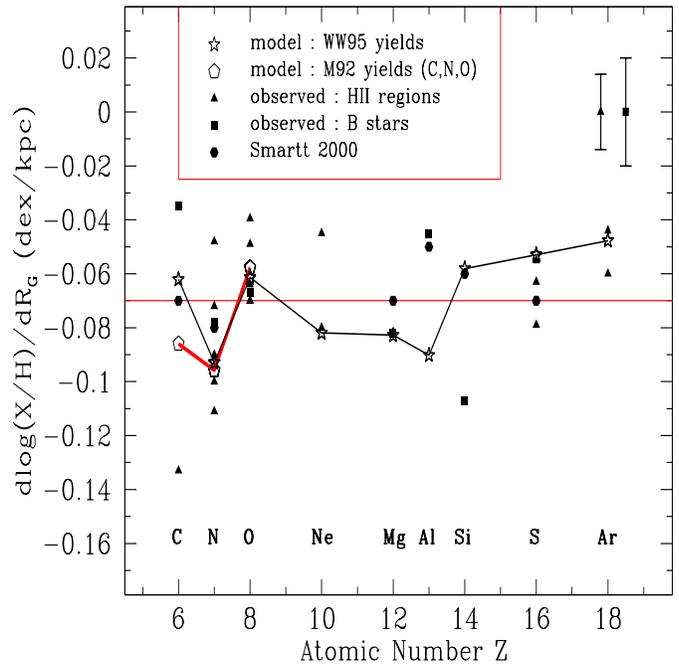,angle=-90,height=9.cm,width=0.5\textwidth}
\caption{\label{} 
Present day abundance gradients in the Milky Way disk: summary of 
models vs. observations. Model results are shown by {\it open symbols}
({\it asterisks}, for the WW95 yields, connected by a {\it solid curve}, 
and {\it pentagons} for the M92 yields,
only for C, N and O). Observations (Table 1 for references) 
are shown by {\it filled symbols}.
Notice the large discrepancy between the 
carbon gradients observed in B-stars and HII regions, 
respectively. An average gradient of $-$0.07$\pm$0.02 \dkp  
is compatible with most of the data and the model results.
Most of the observed abundance gradients
are rather well reproduced by the model, but the  case of Al is
problematic (see text).
}
\end{figure}

{\sc Neon, Sulphur, Argon}: These elements have been observed through their emission
lines in both HII regions and PN of all types. In general, HII region abundances suggest
a radial profile similar to the one of oxygen for all three elements, i.e. a
gradient of $-$0.07 \dkp \ (the flat S profile of Shaver et al. 1983 is an exception
to this general agreement). Data from PNI and PNII support gradients of this
magnitude also for S and Ar, but suggest a smaller gradient for Ne
($-$0.036 \dkp, according to Maciel and Quiroza 1999).

Our model leads to S and Ar abundance gradients similar to the one of O ($\sim -$0.05 
\dkp \ in both cases). In the case of Ne, the (unexcpected) small metallicity
dependence of the WW95 yields (Fig. 1) leads to a steeper profile:
dlog(Ne/H)/dR $\sim$ $-$0.08 \dkp, exactly as in the case of Mg. 

In summary, the observed S and Ar abundance profiles are in agreement with
theoretical expectations, while Ne remains problematic, both  theoretically and
observationally.

{\sc Helium}: For the sake of completeness, we present here the He profiles of
our models. As can be seen in Fig. 7, there is no difference between the profiles
obtained with the WW95 and M92 yields. In both cases a flat He/H profile is 
obtained. However, the lack of IMS yields in our model does not allow to draw 
conclusions about He. 

The observational situation concerning the He abundance profile is not clear. 
The flat profile of the He$^+$/H$^+$ ratio found by Shaver et al. (1983)  was
confirmed by the recent work of Deharveng et al. (2000). However, the true
He/H gradient depends also on the unknown amount of the corresponding neutral 
species. The data on Fig. 6 and 7 are from the work of Gummersbach et al.
(1998) on  B-stars. The error bars are quite large ($\pm$0.3 dex for log(He/H))
and do not allow to draw any conclusion on the existence of an abundance gradient
for He.

Fig. 8 summarizes the discussion of this section, concerning observed and calculated 
abundance profiles in the Milky Way disk. If the values of Smartt (2000) are adopted 
for the C and Si gradients from B-stars, then there is satisfactory agreement between 
the model and observations for all elements but Al (taking into account error bars); 
He and N require another source to account for their absolute abundances, even if the 
gradient has the correct value. But the main ``message'' of Fig. 8 is that homogeneous 
data sets are required for a meaningful comparison between theory and observations.

\subsection{Abundance ratios along the Galactic disk} 

\begin{figure*}
\psfig{file=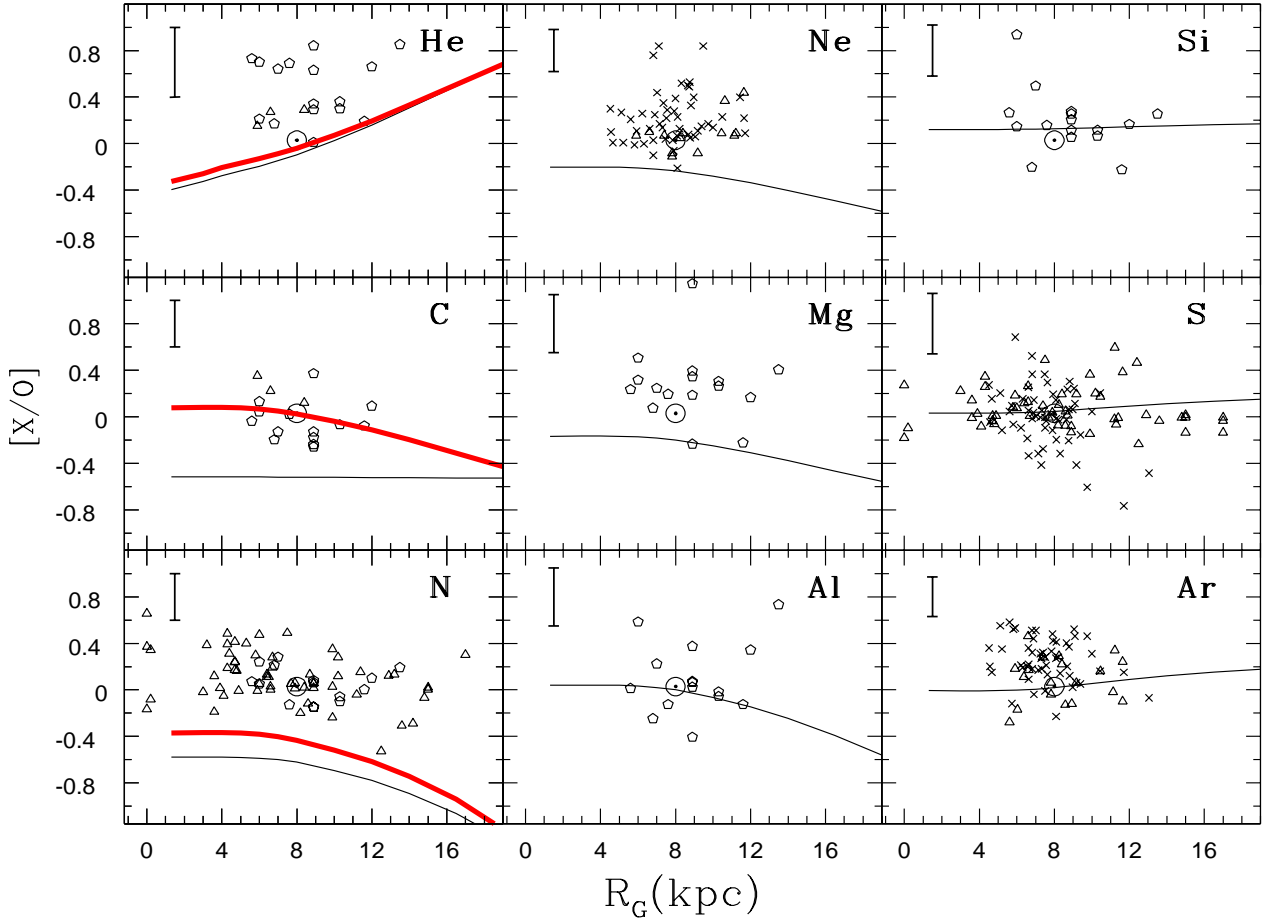,angle=-90,height=13.cm,width=\textwidth}
\caption{\label{} 
Profiles of abundance ratios versus oxygen (normalised to the corresponding
solar ratio: [X/O]=log((X/O)/(X/O)$_{\odot}$)) in the Milky Way disk. 
Observations are shown and compared to 
model results of this work. In all panels, the {\it thin} 
curves indicate results obtained with the WW95 yields, while the {\it thick} 
curves on the {\it left side} panels indicate results obtained with the M92 
yields (leading to larger He/O, C/O and N/O abundance ratios). 
[{\it Note:} The flattening of theoretical profiles of abundance ratios
     in the inner disk
is due to our use of yields for Z=\zs \ even at metallicities larger than solar,
because yields for such metallicities are not available in WW95 and M92].
Data (from references given in Table 1) concern HII regions ({\it triangles}), 
PNI ({\it crosses}) and B-stars ({\it pentagons}). The typical observational 
uncertainties for different element ratios are shown with error bars in the 
upper left corner of each panel. 
Notice that for Ne/O, the best values from type I PNs are 
given by the lower envelope of the data (see discussion in Sect. 4.3).  
}
\end{figure*}

Abundance ratios between metals are, in principle, more reliable tracers of 
the chemical evolution than absolute abundances, since they do not depend on 
the star formation efficiency and they allow to identify if e.g. an element 
has a secondary origin, whether it is produced in long-lived sources 
(low-mass stars or SNIa) or whether it is affected by the ``odd-even'' 
effect, etc. These properties have been widely used as diagnostic tools 
of the history of the halo + solar neighborhood system (se e.g. Pagel 1997). 
In practice, the situation is complicated due to various theoretical and 
observational uncertainties (e.g. Goswami and Prantzos 2000 and references
therein), except for a few trivial cases.

This is also the case for the Milky Way disk, as can be seen in Fig. 9: 
of the nine element abundance ratios to oxygen displayed in the figure,
not a single one shows any clear trend with galactocentric distance.
Although in most cases this is expected on theoretical grounds
(e.g. for Ne, Mg, Si, S, Ar, which are primary elements), for others
(N and, to a smaller extent, He and Al) this is rather surprising. Even
worse is the large scatter obtained for all elements and at all galactocentric
distances. There is a striking difference with F stars in the solar neighborhood, 
which display very small dispersion in their abundance ratios (Edvardsson et al. 
1993). The inhomogeneity of the data sets displayed in Fig. 9 contributes 
certainly to that scatter. However, even in cases where only one tracer is 
involved (B-stars for Mg, Al and Si), the dispersion is rather large:
it is about twice as large as the typical uncertainty of individual abundance 
ratios, estimated here to be $\pm$0.2 dex (see also Henry and Worthey 1999).

A meaningful comparison between theory and observations is difficult in such 
conditions. However, Fig. 9 allows to draw some conclusions:

i) The theoretical abundance ratios of Si/O, S/O and Ar/O are $\sim$ constant 
across the disk, as expected for primaries;  their absolute value is always 
$\sim$ solar and compatible with available observations. The WW95 yields 
reproduce well the current average abundance ratios of those elements.

ii) The theoretically expected ``odd-even'' effect for Al/O is not manifested
in the available observational data, i.e. no trend of Al/O with galactocentric
distance (or metallicity) is observed. The WW95 yields reproduce correctly 
the current local Al/O ratio ($\sim$ solar), and lead to a small dependence on 
metallicity (or radius), because of the ``odd-even'' effect.

iii) The WW95 yields underproduce the observed Ne/O and Mg/O ratios (as anticipated
from Fig. 1); this is also the case for the solar neighborhood, as discussed in 
Goswami and Prantzos (2000). The discrepancy is not large, taken into account 
the various uncertainties of stellar nucleosynthesis calculations (see Prantzos
2000 for a review) as well as in observational data. 
Our Ne/O data come  mainly from PN I, where this ratio is usually determined 
under the assumption that Ne/O = Ne$^{++}$/O$^{++}$. The Ne$^{++}$/O$^{++}$ 
values probably are upper limits to the Ne/O value for central stars hotter 
than 50,000 degrees, due to the presence of charge exchange reaction O$^{++}$ 
+ H$^{0}$ $\Longrightarrow$ O$^{+}$ + H$^{+}$ that allows some O$^{+}$to 
coexist with Ne$^{++}$; this is particularly the case for PNs of type I with 
low density (Peimbert et al. 1995). Therefore the best values from type I PNs 
are given by the lower envelope of the data presented in Fig. 9 (M. Peimbert, private
communication). 
On the other hand, the WW95 yields of Ne and Mg manifest an unexplained 
dependence on metallicity. Put together, these discrepancies point to some 
problems in the WW95 yields of Mg and Al. One possibility is that the extent 
of the C-shell (where both elements are mainly synthesized) is underestimated 
by the Ledoux criterion for convection employed in the WW95 calculations.

iv) The WW95 yields underproduce the local C/O ratio, while the M92 yields of 
massive stars reproduce it fairly well (with no need for a C contribution by 
IMS). However, if WR stars are the main producers of C, one expects this ratio 
to decrease with galactocentric distance (as indicated by the thick curve for
C/O in Fig. 9), and this trend is not seen
in currently available data. On the other hand, if IMS stars produce
the bulk of C as primary, no variation of the C/O ratio with galactocentric 
distance is expected. It should be noted that  gradients of C/O 
(in fact, variations of C/O with O/H) have been observed in extragalactic HII
regions (e.g. Garnett et al. 1999) and are also expected in the case of the Milky
Way disk. Thus, accurate determination of the C/O ratio across  the 
Milky Way disk is crucial in determining the roles of IMS and Wolf-Rayet stars 
in carbon production. [{\it Note:} the flattening of the theoretical C/O 
profile in the inner disk is due to our use of M92 yields for Z = \zs even
when the metallicity is larger than solar; in principle, the C/O ratio should
continue increasing in the inner disk. C and O yields 
for metallicities higher than solar are required for a proper calculation].

v) Both WW95 and M92 yields underproduce the N/O ratio in the disk and lead
to a N/O profile declining with radius. IMS stars are expected to be the main
N producers in the disk. The dispersion of presently available data on N/O 
in the disk does not allow to conclude whether they produce N as primary
(i.e. through hot-bottom burning), or as a secondary.
As with the case of C, observations of N/O in extragalactic HII regions show that
N behaves as secondary at high metallicities (Henry et al. 2000). This
does not seem to be the case in the inner regions of the Milky Way disk, at least
with currently available data.

\begin{figure*}
\psfig{file=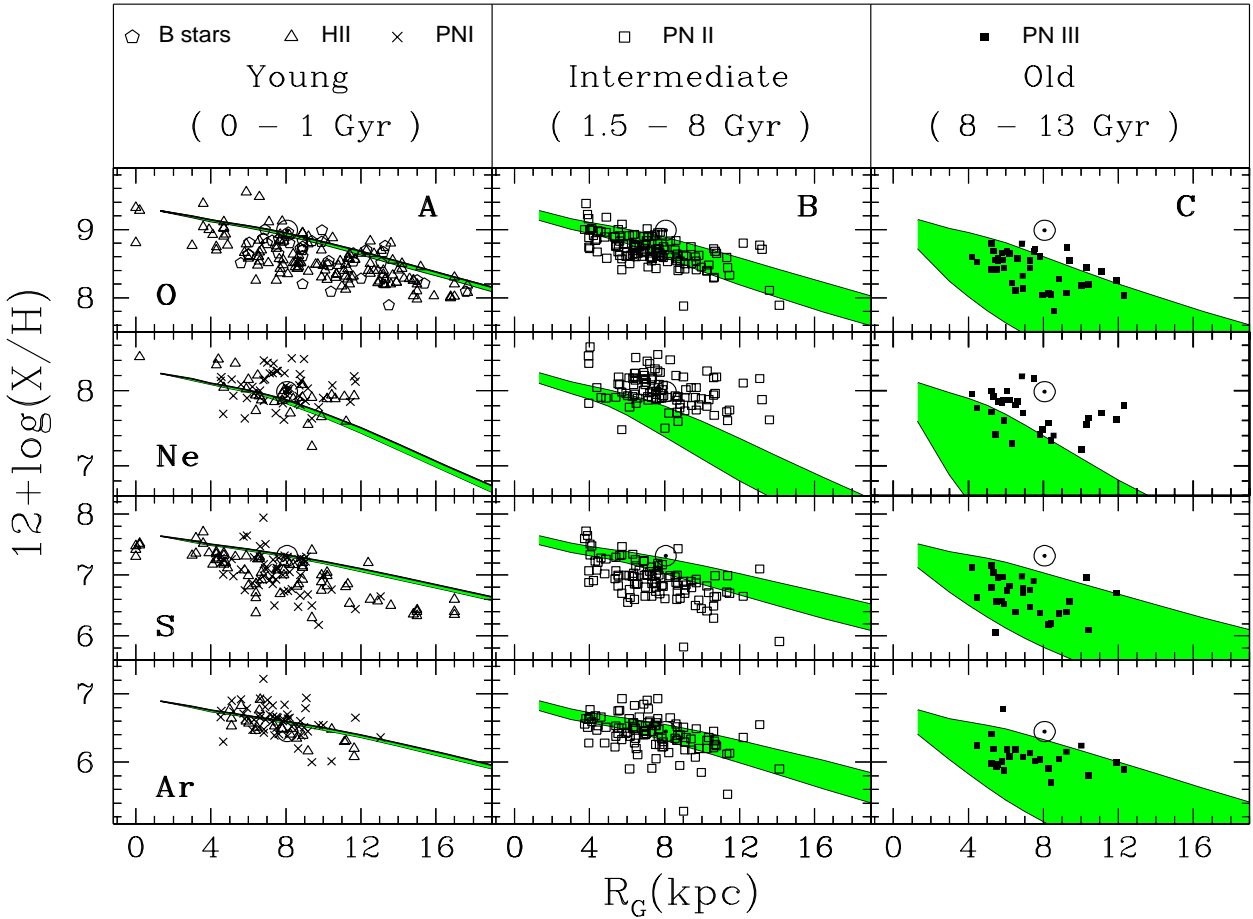,angle=-90,height=13.cm,width=\textwidth}
\caption{\label{} 
Abundance profiles in the Milky Way disk at various epochs according to our 
model and comparison to objects of (hopefully) corresponding ages. From left
to right, profiles are shown for young (column A), intermediate 
(column B) and old age (column C) objects. In Case A belong B-stars, 
HII regions and planetary nebulae of type I; in Case B belong planetary 
nebulae of type II; and in Case C belong planetary nebulae of type III 
(see Sect. 4.4 for a discussion on ages). The corresponding model results 
are shown at time 
T = 12.5$-$13.5 Gyr or age = 0$-$1 Gyr ({\it shaded area} in A), 
T = 5.5$-$12 Gyr or age = 1.5$-$8 Gyr ({\it shaded area} in B) and 
T = 0.5$-$5.5 Gyr or age = 8$-$13 Gyr ({\it shaded area} in C). 
References for data are given in Table 1.
}
\end{figure*}

vi) The observed He/O ratio in B-stars is higher than the corresponding solar
value at all galactocentric distances. This is obviously related to the fact that
the O abundances of those young objects are (somewhat surprisingly) 
systematically lower than solar. Since no satisfactory explanation for that
exists up to now (at least in the framework of conventional chemical evolution
models), we do not expect our results to match the average observed value of He/O.
On the other hand, the B-star data show a flat He/O profile (as a result of the
similar He/H and O/H gradients shown on Fig. 6 and 7). In our calculations,
the amount of He produced by massive stars alone (either with the WW95 or
the M92 yields) is negligible relative to the primordial one of the infalling gas.
Thus, the final He/H profile is flat (Fig. 6 and 7) and the final He/O profile
(Fig. 9) is decreasing in the inner disk; obviously, the contribution of
intermediate mass stars is mandatory to account for He observations.

\subsection{Evolution of abundance gradients} 

As already mentioned in Sect. 4.1, abundances in planetary nebulae of various
types (I, II and III) allow, in principle, to follow the history of the
abundance profile of the Milky Way disk. However, the progenitor masses and 
lifetimes of PNs are not well known. For instance, Allen et al. (1998) assume
that both PNII and PNIII have progenitors of similar mass range (M$<$2 \ms)
and classify them in terms of their kinematical properties: PNII have peculiar 
velocities $v_P<$ 60 km/s and PNIII have $v_P>$ 60 km/s. To account for these 
properties as well as for the relative frequency of the various PN types, 
Allen et al. (1998) propose a ``dynamical'' model for the Milky Way disk, 
simulating orbital diffusion. They find good agreement with observed gradients 
of PN (classified according to that scheme), provided that not all IMS go 
through the PN stage. In that same paper, they also explore the ``conventional'' 
scheme, classifying PN in terms of progenitor mass: 1.3$<$M/\ms$<$8.4 and 
$\tau<$ 3 Gyr for PNI, 0.9$<$M/\ms$<$1.3 and 3$<$ $\tau<$ 9 Gyr for PNII and 
0.8$<$M/\ms$<$0.9 and $\tau>$ 9 Gyr for PNIII.  They find that their model 
fails to reproduce observations of the various PN types in that case. 

In our model we have no information on the kinematical properties of the gas or
the stars. We adopt then the classification scheme of Pasquali and Perinoto (1993) 
for PN, already presented in Sect. 4.1. We are aware that this scheme is not 
necessarily the most appropriate one (uncertainties in ages may be larger than 
stated and kinematics may play a non negligible role), but we adopt it for 
the sake of comparison to our model results. This comparison is presented in 
Fig. 10. It can be seen that:

i) The model describes relatively well the O, Ne, S and Ar abundance gradients
of young objects (left hand panels, column A), as already discussed in the 
previous sections. However, the observed scatter at all galactocentric distances 
is much larger than the range of values given by the model for the last 1 Gyr 
(the shaded region in Fig. 10). It should be noted that the observed scatter 
($\sim$ 0.6 dex) is larger than the corresponding one for young stars in the
solar neighborhood (e.g. Garnett and Kobulnicky 2000). Probably, systematic 
effects and uncertainties in distance estimates contribute largely to the 
observed scatter of ``young'' tracers in Fig. 10.

ii) Model predictions are compatible with observations for O and Ar in PNII, 
as can be seen in the middle panels of Fig. 10 (column B). In fact, there is 
very little difference between the data for PNII and those for PNI. Predictions 
for S are slightly above the data points, while predictions for Ne are clearly 
below the corresponding data. The reason for the latter discrepancy is obviously 
the unexplained underproduction of Ne in the low metallicity yields of WW95, 
already discussed in Sects. 2.2 and 4.3. In the other 3 cases (O, S and Ar) there 
is fairly good agreement between theory and observations concerning all properties 
of the observed abundance profiles (absolute values, gradient, scatter).

iii) Observations show that, on average, abundances in PNIII are lower and present 
a larger dispersion than in PNII. Both features are relatively well reproduced by
our models, as can be seen on the right hand panels in Fig. 10 (column C), with the 
exception of Ne (for the reasons already mentioned in the previous paragraph).

Our models, as some other models of this kind (Moll\'a et al. 1997, Portinari and Chiosi
1999) suggest that abundance gradients are steeper for older objects. 
Other works (e.g. Chiappini et al. 1999) reach an opposite conclusion, although
they are based on the same assumption, namely an ``inside-out'' formation
of the disk; clearly, the adopted time-scales for star formation and infall play
an important role in this discrepancy. Although our 
model results are compatible with available data on PN, the large scatter in those 
data does not allow to conclude on the temporal variation of the gradients
(see Maciel and Quireza 1999). However, 
our model makes a new, and perhaps testable, prediction: if the observed abundance 
scatter at a given galactocentric distance is to be attributed, at least partially, 
to intrinsic age differences between the tracers (e.g. PN)
then we expect that scatter to be smaller in the inner 
disk than in the outer one (as indicated by the extent of the shaded areas in Fig. 10).

\section{Summary}

Based on the successful chemical evolution model for the Milky Way disk
developped by BP99, we calculated the corresponding abundance profile of elements 
up to the Fe-peak. For that purpose we used the metallicity dependent yields
of WW95 for massive stars evolving at constant mass, along with those
of Iwamoto et al. (1999) for SNIa. We also explored  the yields of M92 for 
massive mass losing stars, concerning the elements He, C, N and O. We
deliberately  ignored yields from intermediate mass stars (IMS),  
our purpose being to check to what extent such stars are indeed required
to explain observed abundance patterns. However, we did take into account ejecta
from IMS, assuming that their net yield is zero for all elements; as
explained in Goswami and Prantzos (2000), this procedure is crucial to ensure that
the absolute metal abundances, expressed as X/H, are correctly evaluated.

We compiled a large sample of observational data, concerning abundance
profiles of He, C, N, O, Ne, Mg, Al, Si, S and Ar in the Milky Way disk (Table 1).
Most of the tracers are young objects (B-stars, HII regions, PNI), while
PNII and PNIII abundances trace earlier stages of the Galaxy evolution
(but with considerable uncertainties in the corresponding ages). Contrary
to other authors, we did not
consider data on Fe from open clusters, since the situation is rather uncertain
at  present (see Sect. 4.1). We simply note that our prescription for the rate
of SNIa (major Fe producers), leads to a Fe abundance profile slightly steeper 
than the one of O in the disk (Fig. 4 and 5).

The main results of the model may be summarized as follows:

(i) We obtain abundance gradients for all elements from He to Zn. For primary 
elements, like O, the theoretical value of the abundance gradient at the present
epoch (T = 13.5 Gyr) is: dlog(X/H)/dR $\sim-$0.06 \dkp. This value is compatible
with most observational data concerning young objects in the disk (see Fig. 8 and 
Table 1). 
We note, however, the ``puzzling'' conclusion of Deharveng et al. (2000), pointing
to an oxygen gradient smaller by about 40\% 
than the commonly accepted value;
if their conclusion is confirmed, our model parameters 
(the ratio of star formation to infall timescale as a function of galactocentric 
radius) should be appropriately modified. 
For secondary or ``odd-Z'' elements (like N and Al, respectively)
we obtain slightly larger gradients; this is also the case for C when the M92 
yields are used. Such values are marginally compatible with available 
observations. We find that the M92 yields can account for the totality of
carbon production across the disk with no need for a contribution by IMS,
in agreement with other recent studies (Prantzos et al. 1994, Carigi 1994,
Gustafsson et al. 1999, Henry et al. 2000a, 2000b). On the contrary, the majority 
of N is produced by another source, most probably IMS.

(ii) Current observations show no trend of the abundance ratio X/O with
galactocentric distance (or metallicity), 
for any element X. This is unexpected in the case 
of N and C, which do show such a trend in extragalactic HII regions where
their abundance ratio to oxygen increases with metallicity (e.g. Henry and 
Worthey 1999). Our model shows a flat profile of S/O, Si/O and Ar/O
(in agreement with observations) and  a slowly decreasing ratio of Ne/O, Mg/O
and Al/O with galactocentric distance (due to a ``problematic'' or 
overestimated metallicity dependence of the WW95 yields for those elements).
They also show that C/O should decrease with galactocentric distance
in the case of M92 yields (but this should not be the case  if C is mainly 
produced as a primary by IMS). Clearly, precise observations of the
C/O ratio across the disk are required in order to decide about its main 
production site.
Similar conclusions hold for N: the M92 yields cannot account for its
abundance profile, but is not clear whether primary or secondary production in
IMS is the dominant mechanism. Observations of extragalactic HII regions 
suggest that the latter mechanism dominates in
high metallicity regions (Henry et al. 2000b);
if this is true, then an important N/O gradient should be observed in the
Milky Way disk.

(iii) The evolution of abundance gradients provides a strong constraint for
Galactic chemical evolution models. Despite a considerable amount of
observational and theoretical work (e.g. Henry and Worthey 1999; Tosi 2000
and references
therein), the question has not yet been  settled.
Observations of planetary  nebulae (PN) of different ages could, 
in principle, help in that respect (Pasquali and Perinoto 1993, Maciel 
and Quireza 1999), but the ages and 
distances attributed to PN are not sufficiently well known at present. 
Our model predicts a steady
flattening of the gradients with time, due to the adopted ``inside-out''
formation scheme for the disk. Such an evolution is also found in other
works (e.g. Moll\'a et al. 1997, Portinari and Chiosi 1999) using
similar assumptions. Our model makes also a
testable prediction: the abundance scatter must be  smaller in the 
inner disk than in the outer regions, 
{\it if } the observed dispersion between PN abundances
at a given galactocentric distance is due
to their intrinsic age differences.

In summary, we have shown that current massive star yields, combined with
the BP99 chemical evolution model, reproduce fairly well most of the
observed abundance gradients in the Milky Way disk. 
The assumptions of the model affect mostly points (i) and (iii) above, while
the adopted yields affect point (ii).
Some problems remain
with C and N, and (to a smaller extent) Al. However, the major question
of such studies, namely the evolution of abundance gradients, is far from
being settled yet; a much more precise determination of ages, distances and
abundances of planetary nebulae will be required for that.

\acknowledgements J.L. Hou acknowledges the warm hospitality of the 
IAP(Paris, France). We are grateful to the referee, Dr. M. Peimbert, for his 
suggestions that improved the paper.
This work was made possible 
thanks to the support of China Scholarship Council(CSC) and National 
Natural Sciences Foundation of China.

\def\apj{ApJ}
\def\apjl{ApJL}
\def\apjs{ApJS}
\def\aj{AJ}
\def\aap{A\&A}
\def\araa{ARA\&A}
\def\aapss{A\&AS}
\def\mnras{MNRAS}
\def\nature{Nature}
\def\apss{Ap\&SS}
\def\pasp{PASP}

{}

\end{document}